\documentclass[conference]{IEEEtran}

\usepackage{makeidx}  
\usepackage{multicol, blindtext}
\usepackage{lipsum}
\usepackage{colortbl}
\definecolor{Gray}{gray}{0.9}
\usepackage{hhline}

\usepackage{graphicx}
\usepackage[ruled,vlined,algonl]{algorithm2e}

\usepackage{amsmath}

\usepackage{multirow}
\usepackage{amssymb,latexsym}
\usepackage{cite}
\usepackage{color}
\usepackage{listings}
\usepackage[tight]{subfigure}
\usepackage{gensymb}
\usepackage{bbding}
\usepackage{listing}
\usepackage{flushend}
\usepackage{url}

\begin{document}

\title{A Dynamic Overlay Supporting Just-In-Time Assembly to Construct Customized Hardware Accelerators}

\author{
\IEEEauthorblockN{ Zeyad Aklah, Sen Ma, David Andrews}
\IEEEauthorblockA{Department of Computer Science and Computer Engineering\\
University of Arkansas\\
Fayetteville, AR 72701, USA\\
Email: \{zaklah, senma, dandrews\}@uark.edu}
}
\maketitle
\vspace{-4em}
\begin{abstract}
Barriers that prevent programmers from using FPGAs include the need to work within vendor specific CAD tools, knowledge of hardware
programming models, and the requirement to pass each design through synthesis, place and route. In this work, a dynamic overlay is designed to support Just-In-Time assembly by composing hardware operators to construct full accelerators. The hardware operators are pre-synthesized bitstreams and can be downloaded to Partially Reconfigurable(PR) regions at runtime.

\end{abstract}

\section{Introduction}\label{sec:introduction}

Despite the significant advancements that have been made in High Level Synthesis (HLS), the reconfigurable
computing community has failed at getting programmers to use Field Programmable Gate Arrays (FPGAs). Existing barriers that prevent
programmers from using FPGAs include the need to work within vendor specific CAD tools, knowledge of hardware
programming models, and the requirement to pass each design through synthesis, place and route.

Intermediate Fabrics in the form of overlays have been proposed to close the semantic gap between high level languages and
low level hardware.  In their simplest form overlays form virtual \emph{programmable} blocks that can be implemented on top of the FPGAs lookup tables and flip flops~\cite{Coole:2015:FCCM}.  A wide range of overlays have been proposed, from programmable vector processors, complete CGRA type structures~\cite{Cong:2014:FPD} and
programmable interconnect networks~\cite{matrix1}.

In our recent work, we have been exploring how a Just-in-Time (JIT) approach can be used with an overlay to enable programmers to assemble gates through compilation instead of synthesis~\cite{Sen:2015:FPL}.
JIT techniques have been previously investigated for  programming predefined overlay components such as ALUs as well as moving bitstreams into and out of partial reconfiguration regions.

Our work extends these earlier approaches by embedding partial reconfiguration regions within a programmable interconnect overlay.  Combining the use of PR regions
within an overlay provides the potentials advantage of customization and increased resource utilization when compared to a more general purpose
fully static CGRA type structures.  Programmers access  libraries of pre-synthesized parallel patterns  such as map, reduce, foreach, and filter then can be assembled within the FPGA
by a run time interpreter.  This flow allows programmers to
create unique hardware accelerators by composing and compiling symbolic links to different numbers, types, and organizations of library patterns within their source code.
The source code, with symbolic links, is compiled into a series of interpreter instructions executed by the run time system on how to assemble custom bitstream versions of the
programming patterns into the PR regions and set the programmable connections of the communication overlay.
While the basic approach does remove the use of CAD tools, synthesis, place and route from the programmers design path, our current implementation incurs
the following two limitations:

\begin{itemize}
\item \emph{All variants of programming patterns must be synthesized.}
\item \emph{Cannot compose simple conditionals with pre-synthesized programming patterns.}
\end{itemize}

We are now exploring how the functionality of the overlay can be modified to remove these restrictions.

\section{Overlay Architecture}\label{sec:overlay}
\begin{figure}[t]
\centering
\includegraphics[height=.35\textwidth]{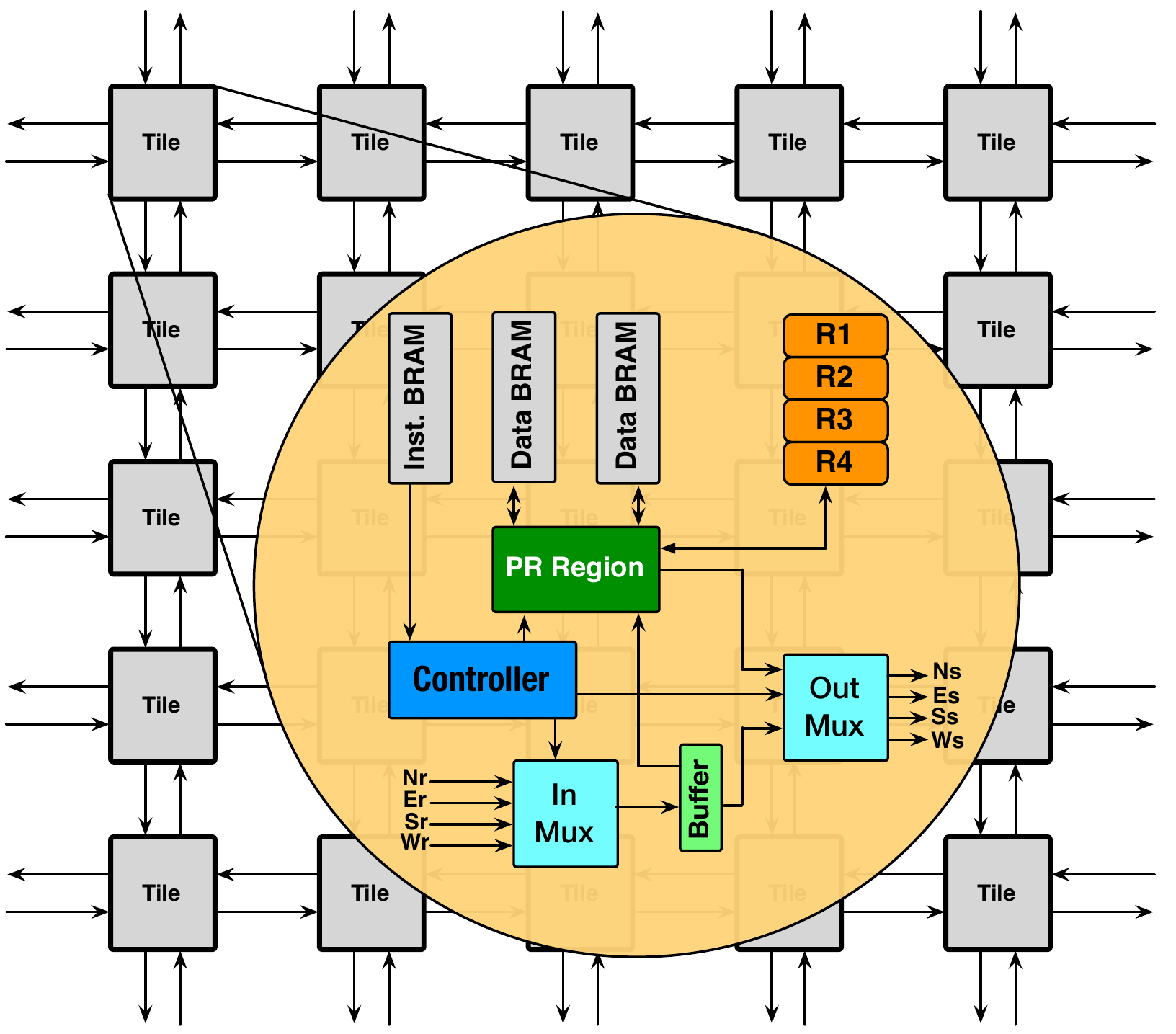}
\caption{\label{fig:structure} Reconfigurable dynamic overlay.}
\end{figure}

Figure~\ref{fig:structure} shows our new overlay that contains PR tiles connected in a 2D mesh like interconnect. Our original overlay is described in~\cite{Sen:2015:FPL} and contained only PR regions with a programmable N-E-S-W interconnect. Page limitations prevent us from providing a detailed explanation of the original overlay.  The following paragraphs describe the new addition and modification we have included to address the two limitations listed above.
In both overlays the number of tiles can be set based on the resource capabilities of each FPGA.
In the new version, each tile has an additional set of registers and three BRAMs; one for instructions and two for data. In the original overlay only the
border tiles had BRAMs for data, and no BRAM for instructions.  The functionality of the both overlays' interconnect is configured by the controller based on the data flow dependencies between the instructions assigned to each tile.  The interconnect allows each tile to consume or bypass (for branching) data into and out of the tile.

The new controller currently interprets 42 different instructions (interconnect: 22 instructions, branching:
6 instructions, vector operations: 2 instructions, Memory \& Register operations: 12 instructions).  As resource usage
varies between different operators we are studying the benefits and drawbacks of setting PR regions within a Tile to different sizes.  In our current systems
we size 1/4 of the PR
regions to contain 8 DSPs, 964 FF, and 1228 LUTs.  This supports our larger operators such as $sqrtf, sin, cos, log$.  The remainder are set to 4 DSPs, 156 FF, and 270
LUTs. This particular configuration is more for convenience than optimality, due to the current layout of physical resources within our FPGAs (Xilinx series)  and PR flow
constraints.  We are using this configuration to study how such non-uniform organizations can reduce the internal fragmentation within the PR regions versus flexibility of mapping and performance.

We are currently focused on how the use of such a dynamic overlay can support conditional branching, reduce the total number of bitstreams that need to be produced versus performance and resource utilization.   Our studies include understanding if the use of smaller tiles will enable higher gate densities by allowing only active operators to be resident within the overlay.  We are evaluating if this will allow more active tiles to be packed into a given unit area compared to a static approach.

Our overlay currently supports conditional branching with speculation through an ability to dynamically map operators and set the interconnect at run time.
We want to understand how this configuration supports better performance in the presence of conditional branching by allowing if-then-else operators to be placed within contiguous tiles.  Dynamic mapping in general has the potential to reduce latencies that arise if operators are contained within non-contiguous tiles.   This mapping may facilitate better pipelining.   This also has the potential to simplify the runtime scheduler.

\begin{figure}[t]
\centering
\includegraphics[height=.15\textwidth]{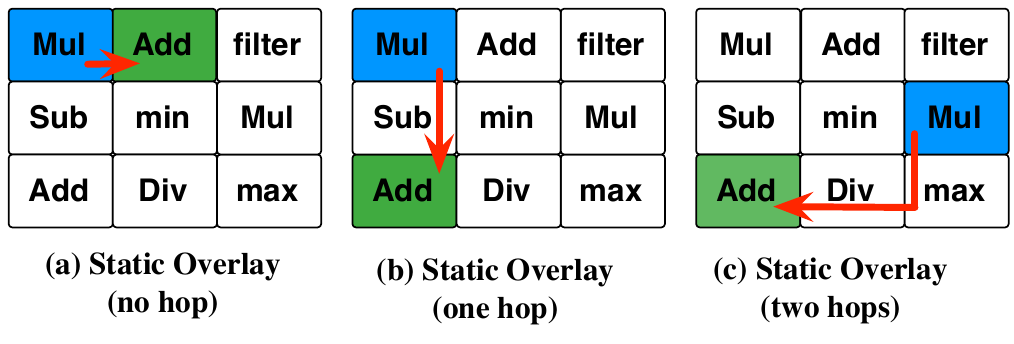}
\caption{\label{fig:static_overlay} Mapping VMUL\&Reduce patterns into static overlay using three scenarios.}
\end{figure}

\begin{figure}[t]
\centering
\includegraphics[height=.23\textwidth]{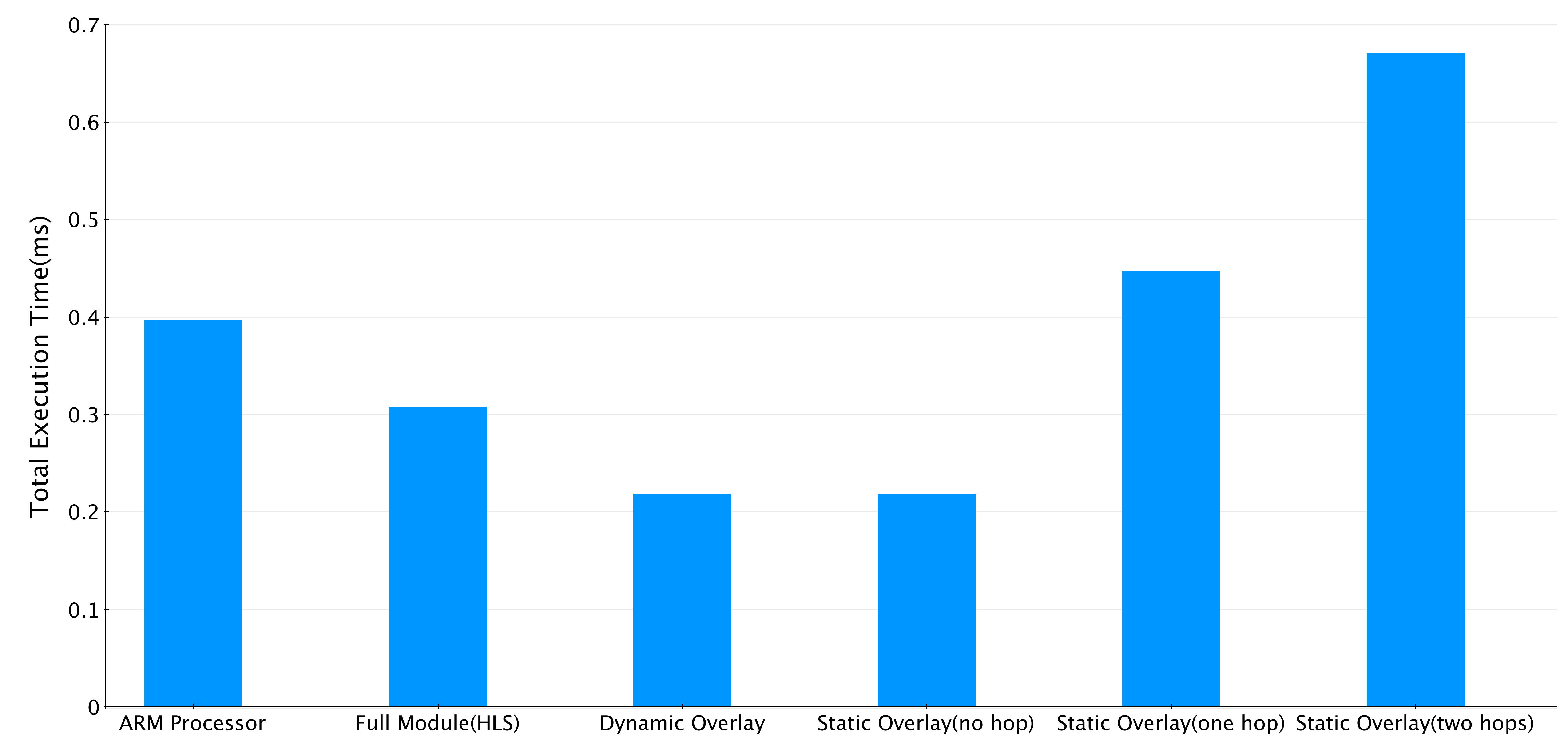}
\caption{\label{fig:performance} Performance}
\end{figure}
\section{Preliminary Results}
The overlays used in this study were created through an automated script that produced TCL scripts that were input into the CAD tools for synthesis.
All systems were built on a Virtex7 and synthesized using the Vivado 15.3 tools. Vector-Multiply(VMUL) and Reduce ($sum=\sum \vec{A} \times \vec{B} $) patterns were created and used in our initial experimentation.  Two configurable 3$\times$3 overlays (static and dynamic) were built.
Figure~\ref{fig:static_overlay} shows how the operators are organized in the static overlay.  This specific organization was defined to allow us to measure the penalty
of having non contiguous operators.  The VMUL and Reduce patterns needed at least one multiplier and one adder.  Three scheduling scenarios were tested as
shown on Figure~\ref{fig:static_overlay}.

Figure~\ref{fig:performance} shows the total execution time of VMUL and Reduce in milliseconds using five different hardware targets; the data size was set to 16 KBytes. The total
execution time includes data transfer and execution time. The dynamic overlay shows better performance since operators are always contiguous and pipelined.  The only penalty of
the dynamic overlay is the PR overhead which is around (1.250 ms) in this experiment.  As this time would only be incurred at startup or initial configuration, it has not been
included in the graph.  The performance of the static overlay decreases as the number of pass through tiles increases.
The fully custom module used for comparative purposes was designed using Vivado HLS.  The design was not optimized to reflect a closer performance to designs built with HLS by non hardware experts. The Zedboard was used to test the performance on 660 MHz ARM processor.

\bibliographystyle{IEEEtran}
\bibliography{IEEEabrv,OLAF}

\begin{thebibliography}{1}
\providecommand{\url}[1]{#1}
\csname url@samestyle\endcsname
\providecommand{\newblock}{\relax}
\providecommand{\bibinfo}[2]{#2}
\providecommand{\BIBentrySTDinterwordspacing}{\spaceskip=0pt\relax}
\providecommand{\BIBentryALTinterwordstretchfactor}{4}
\providecommand{\BIBentryALTinterwordspacing}{\spaceskip=\fontdimen2\font plus
\BIBentryALTinterwordstretchfactor\fontdimen3\font minus
  \fontdimen4\font\relax}
\providecommand{\BIBforeignlanguage}[2]{{%
\expandafter\ifx\csname l@#1\endcsname\relax
\typeout{** WARNING: IEEEtran.bst: No hyphenation pattern has been}%
\typeout{** loaded for the language `#1'. Using the pattern for}%
\typeout{** the default language instead.}%
\else
\language=\csname l@#1\endcsname
\fi
#2}}
\providecommand{\BIBdecl}{\relax}
\BIBdecl

\bibitem{Coole:2015:FCCM}
J.~Coole and G.~Stitt, ``Adjustable-cost overlays for runtime compilation,'' in
  \emph{Field-Programmable Custom Computing Machines (FCCM), 2015 IEEE 23rd
  Annual International Symposium on}, May 2015, pp. 21--24.

\bibitem{Cong:2014:FPD}
\BIBentryALTinterwordspacing
J.~Cong, H.~Huang, C.~Ma, B.~Xiao, and P.~Zhou, ``A fully pipelined and
  dynamically composable architecture of cgra,'' in \emph{Proceedings of the
  2014 IEEE 22Nd International Symposium on Field-Programmable Custom Computing
  Machines}, ser. FCCM '14.\hskip 1em plus 0.5em minus 0.4em\relax Washington,
  DC, USA: IEEE Computer Society, 2014, pp. 9--16. [Online]. Available:
  \url{http://dx.doi.org/10.1109/.10}
\BIBentrySTDinterwordspacing

\bibitem{matrix1}
\BIBentryALTinterwordspacing
\emph{{MATRIX: a reconfigurable computing architecture with configurable
  instruction distribution and deployable resources}}, 1996. [Online].
  Available: \url{http://ieeexplore.ieee.org/xpls/abs\_all.jsp?arnumber=564808}
\BIBentrySTDinterwordspacing

\bibitem{Sen:2015:FPL}
S.~Ma, Z.~Aklah, and D.~Andrews, ``A run time interpretation approach for
  creating custom accelerators,'' in \emph{Field Programmable Logic and
  Applications (FPL), 2015 25th International Conference on}, Sept 2015, pp.
  1--4.

\end{thebibliography}
\vfill
\end{document}